\documentclass{article}
\usepackage[utf8]{inputenc}
\usepackage{graphicx}
\usepackage{caption}
\usepackage{subcaption}
\usepackage{hyperref}
\usepackage[
 a4paper, 
 total={170mm,240mm},
 left=20mm,
 top=20mm,]{geometry}
\usepackage[
backend=biber,
style=authortitle,
]{biblatex}
\usepackage{multirow}
\usepackage{mathtools}

\graphicspath{ {./images/} }
\hypersetup{
    colorlinks=true,
    linkcolor=blue,
    filecolor=blue,      
    urlcolor=blue,
    citecolor=blue,
    pdftitle={Overleaf Example},
    pdfpagemode=FullScreen,
    }

\renewenvironment{abstract}
 {\small
  \begin{center}
  \bfseries \abstractname\vspace{-.5em}\vspace{0pt}
  \end{center}
  \list{}{
    \setlength{\leftmargin}{.8cm}%
    \setlength{\rightmargin}{\leftmargin}%
  }%
  \item\relax}
 {\endlist}

\title{%
\textbf{The Economics of the DeLend Project \\
\Large Agent-based Simulations} \vspace{0.5cm}}
\author{%
    Frederico Dutilh Novaes, PhD \\
    {\small fred@delend.finance}
    \and
    Gabriel de Abreu Madeira, PhD \\
    {\small gmadeira@usp.br}
    \and
    Aurimar Cerqueira \\
    {\small harry@delend.finance}
}

\date{\vspace{0.5cm} \textbf{February 2023}}

\begin{document}

\maketitle

\begin{abstract}
This paper presents our methodology to simulate the behavior of the DeLend Platform. Such simulations are important to verify if the system is able to connect the different sets of agents linked to the platform in a functional manner. They also provide inputs to guide the choices of operational parameters, such as the platform spread, and strategies by DeLend, since they estimate how the key variables of interest respond to different policies. We discuss the methodology and provide examples meant to clarify the approach and to how we intend to use the tool in practice -- they should not be interpreted as representative of real life scenarios.\\

\end{abstract}

\section{Introduction}
Complex collective behaviors can, sometimes, be modeled by decomposing systems into a set of \emph{agents} that interact between themselves following relatively simple rules/interactions. Simulations built on such modeled agents are called agent-based simulations. Our agent-based simulation tool is presented, which is designed to simulate the behavior of the DeLend platform, and illustrative examples of such simulations are provided. The numerical exercises provided build on the market design presented on \cite{delendwp1}. DeLend connects, through its  platform, different sets of agents: borrowers, investors and guarantors. The parameters governing the platform must be set so that there is coherency among the actions of such actors. Resources anticipated for borrowers should be similar to the amount provided by investors. Guarantors must have incentives to truly facilitate the borrowing process \footnote{ Incentives constraints are crucial for guarantors to be interested in endorsing loans and for such endorsement to be worthwhile for borrowers and investors. See for instance \cite{townsend2020distributed} and \cite{salanie2005economics}}. Simulations can be used to verify if such properties are achievable with the platform design proposed. Additionally, they provide an estimate of the gains obtained by each agent connected to the platform. Finally, simulations reveal impacts on amounts lent, prices and return to investors of different parameters under multiple scenarios, thus guiding the choices of operational parameters and policies by DeLend. 

We start by briefly introducing the agents and their interactions\footnote{Although a simplified representation of reality, the specification of these agents still involves some level o math, and their implementation must be thoroughly tested. This more in-depth discussion has not been included in this main text, as it is meant to be an overview.}, including a discussion on the important topic of (loan) demand modeling. Next,  we simulate somewhat realistic examples, where all classes of agents are present, and obtain the optimal spreads on different scenarios, as an example usage of the tool. The results indicate that the system is capable of channeling the resources from investors to borrowers with a reasonable return for investors and relative stability of the liquidity pool. They also suggest that a good engine for rating loans is crucial, and that guarantors are an important tool to improve the performance of the system whenever the rating engine is not sufficiently accurate. We then conclude and discuss next steps.

\section{Agent-based simulation}
Agent-based simulations require a characterization of the behavior of agents and rules of interaction amongst such agents. The set of agents and their interaction rules are described below. 

\subsection{Agents}

\begin{figure}[h!]
    \centering
    \includegraphics[width=0.6\textwidth]{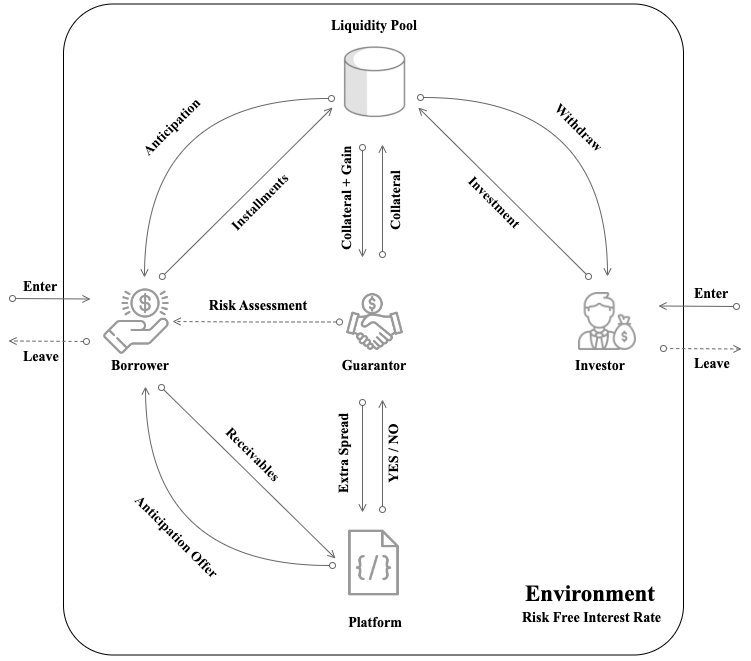}
    \caption{Agents and their interactions. We also include environmental parameters, so far we have only considered a \emph{risk free interest rate}, that acts as a base rate.}
    \label{fig:agents}
\end{figure}

Three classes of agents drive the simulations: investors, borrowers and guarantors. Those are the agents that will connect through the DeLend platform. Two classes of investors are modeled: (i) seed investors, which provide liquid assets to the platform initially; and (ii) regular investors, which provide liquid assets for lending over time. The number of those agents usually varies over time -- agents can enter and leave the system. The behavior of agents depends on preference parameters and information available to them.

The Platform is the engine that computes the anticipation offers for each potential borrower -- it may also refuse to offer an anticipation, if it considers the borrower to be too risky. The anticipation is computed considering four main parameters: (i) a risk evaluation (quantified as a default probability); (ii) the environment's risk free interest rate; (iii) a spread, which stipulates extra returns per loan (when compared to the risk-free rate) (iv) guarantors contributions. Section \ref{sec:borrower_platform} discusses the basic setup, without guarantors. If guarantors are involved, the offer will take into account the guarantor's stake and their (guarantors) expected gains (see section \ref{sec:guarantors}).

The Liquidity Pool manages the cash flow, by transferring and receiving resources from borrowers and investors. Given the dynamical nature of the system, there may be periods where there are not enough resources to anticipate or to fully pay an investor (in which case, the investor will only partially withdraw, since we have not yet implemented a secondary market yet). Thus, the management of the liquidity pool is a key piece of the DeLend system.

Figure \ref{fig:agents} depicts the agents and their interactions. Since the interactions between borrowers and guarantors with the liquidity pool are very simple, being simply a cash transfer, we will only focus on describing the interactions between borrowers and the platform, with and without guarantors, and between investors and the liquidity pool\footnote{In reality, investors -- and borrowers, and guarantors -- interact only with the platform, the latter managing the liquidity pool. Here we are representing as if they interacted directly with the liquidity pool in order to facilitate the representation of the different cash flows.}.

\subsection{Interactions between agents}
The following sections describe the interaction between agents. Investors and guarantors provide funds, while borrowers absorb such resources. The interactions presented are just a subset of the possible options given the structure of the DeLend system. Other arrangements may be analysed in a future work. 

\begin{figure}[h!]
    \centering
    \includegraphics[width=0.7\textwidth]{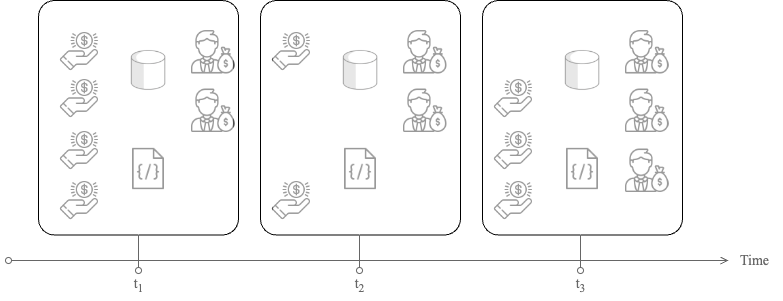}
    \caption{The time evolution is discrete, where each step can be constructed to represent a day, a month, a year, or whatever convenient unit of time. At each time step, agents interact: borrowers can ask for anticipations or pay installments, investors may decide to invest or withdraw (or do nothing), and so on. We have also represented the fact that borrowers and investors may enter/quit the environment, thus their number may vary.}
    \label{fig:time_evolution}
\end{figure}

To actually perform a simulation means to let the system evolve in time. As represented in Figure \ref{fig:time_evolution}, the time evolution is discrete, where, at each time step, agents may interact (take actions) according to the behaviours described bellow.

\subsubsection{Borrower $\longleftrightarrow$ Platform: Anticipation without guarantors}
\label{sec:borrower_platform}

As depicted in Figure \ref{fig:agents}, borrowers may trade a set of receivables for an anticipation. The platform's job is to make an anticipation offer based on the set of receivables (amount and number of installments) and its risk evaluation. If the borrower accepts the anticipation offer, the anticipation value is transferred from the liquidity pool to the borrower, in return for the installments. Over time, the liquidity pool then receives the installments\footnote{For now we have only consider full payment of each installment.}, except for in the event of a default. We now (qualitatively) discuss our approach.

Given a set of $N$ receivables (e.g. installments of a sale), a merchant may want to trade these receivables for cash anticipation. Here, we consider a scenario without guarantors, where, in order to make an anticipation offer, the platform considers two main factors that impact the anticipation discount: (i) the default risk discount (that depends on the estimated default probability we refer as $p$); and (ii) the present value discount (a combination of the risk free interest rate $r$, the platform's spread $s$, and the number of installments $N$).

\begin{figure}[h!]
    \centering
    \includegraphics[width=0.6\textwidth]{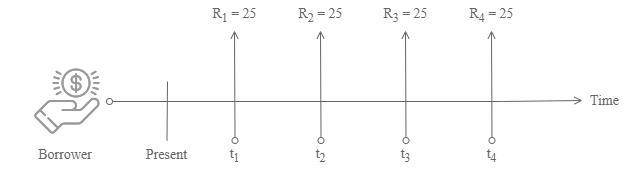}
    \caption{Cash flow of 4 receivables (installments) totaling 100. In general, installments need not have equal values, nor be equally spaced, but often are. Typically, the time periods $t_i$ represent months.}
    \label{fig:baisc_inst_cflow}
\end{figure}

\begin{figure}[h!]
    \centering
    \includegraphics[width=0.6\textwidth]{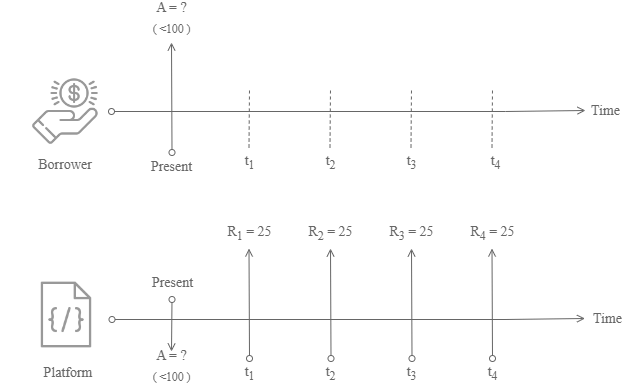}
    \caption{Cash flow if the trade between the borrower and the platform occurs: the platform pays the borrower an amount $A < 100$ for the set of future receivables. Being future payments, there will be a certain level of default to be expected, thus impacting the anticipation discount.}
    \label{fig:baisc_inst_cflow_w_ant}
\end{figure}

As an example of cash flow, suppose a borrower is willing to negotiate a set of $N=4$ (equally spaced) receivables totaling 100 (4 installments of 25), as shown in Figure \ref{fig:baisc_inst_cflow} -- where upward arrows indicate cash inflow, and downward arrows cash outflows. If the platform makes an anticipation offer for the set of receivables and the borrower accepts it, the platform pays an amount A to the borrower in return for the receivables (Figure \ref{fig:baisc_inst_cflow_w_ant}).

The challenge for the platform is to compute the amount $A$ to anticipate, and the loan's return will be directly related to the discount offered: the bigger the discounts, the higher the returns for investors. However, if discounts are too greedy, lower volumes will be captured, since borrowers are more likely to reject anticipation offers, resulting in sub-optimal allocation of the liquidity pool. Also, if risk discounts are not properly computed, returns may become negative and/or lead to adverse selection.

\begin{figure}[h!]
    \centering
    \includegraphics[width=0.8\textwidth]{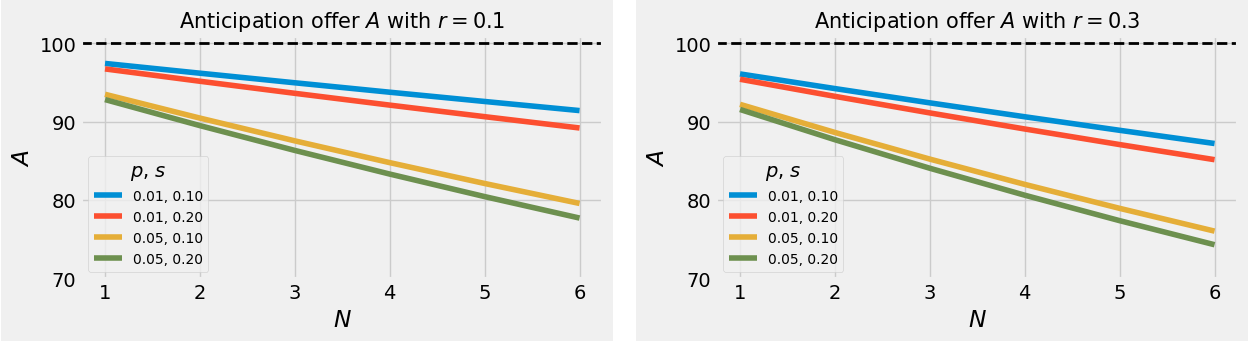}
    \caption{Anticipation values for a set of receivables that sum 100, as a function of the number of installments $N$, the platform's spread $s$, its estimated default probability $p$ and for a base rate $r = 0.1$ (left), and $r = 0.3$ (right). When guarantors are not involved, these are the parameters that we have included to compute the anticipation offer.}
    \label{fig:plot_example_A}
\end{figure}

Figure \ref{fig:plot_example_A} shows how the anticipation value changes as a function of the different parameters we have included in our anticipation formula in this context. As it can be noted, the estimated default probability $p$ significantly impacts the value of the anticipation offer. As long as the platform's risk assessment is accurately estimating these values, these would be the proper discounts in order to have a healthy operation. However, if, for whatever reason, the rating engine is not properly calibrated, guarantors may come to help, as we discuss in the next section.

\subsubsection{Borrower $\longleftrightarrow$ Guarantor $\longleftrightarrow$ Platform: Anticipation with guarantors} \label{sec:guarantors}

Guarantors can contribute in different ways to the dynamics of the system. In particular, when they are able to provide more accurate risk assessments regarding particular borrowers, their role can be beneficial to all players: it enables lending by improving anticipation offers (as we'll discuss below) while also being profitable to the guarantor. 

As shown in Figure \ref{fig:agents}, guarantors participate in the trade between a borrower and a platform: they have their own risk assessment and they ask for an extra spread in return for their investment (collateral). The platform then decides if it is worth involving the guarantor\footnote{The simplest criteria, which is the one we have used, is to establish a threshold of improvement for anticipation -- e.g. 10\% better than without guarantors. However, more elaborate criteria may be needed, for example in order to avoid possible collusion between borrowers and guarantors.}.

The dynamics  works as follows. Guarantors must first deposit a collateral $V_c$. If the borrower defaults, the guarantor loses the collateral, being an investment with higher risks than for standard investors. Thus, the loan's fund return alone, usually, would not be enough as an incentive to attract guarantors, and the platform must also offer an extra gain, characterized by the guarantor's spread $s_g$. Guarantors also have an estimate for the borrower's default probability\footnote{Indirectly acting in a similar way as a \emph{peer prediction index}, \cite{Townsend2019DistributedL}}, in order to compute expected losses. The following elements characterize a guarantor: (i) $s_g$ the guarantor's extra spread; (ii) $V_c$ the guarantor's collateral amount; and (iii) $p_g$ the guarantor's default probability estimate (its risk assessment).

From the platform's perspective, since we have not (yet) included a (extra) financial reward coming from the partnership with the guarantor, the advantage comes from being able to offer better anticipation deals to the borrowers, increasing the acceptance probability. From the guarantor's perspective the reward is profitability.

\begin{figure}[h!]
    \centering
    \includegraphics[width=0.4\textwidth]{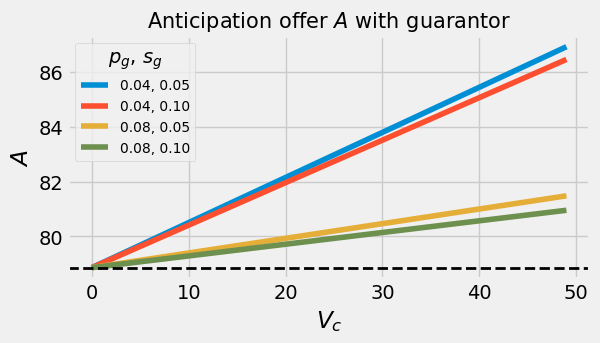}
    \caption{Anticipation offer for a set of receivables that sum 100, with $N = 3$, $s = 0.1$ and $r = 0.1$. The platform estimates the default probability to be $p = 0.1$, while the guarantor's default probability estimate is $p_g$ ($s_g$ is the guarantor's extra spread and $V_c$ is the staked amount). The black, dashed line is the anticipation offer without guarantor. The difference in the default probability estimates is the main factor that impacts the anticipation offer. For this reason, when guarantors are able to provide a better risk assessment, their involvement can significantly improve the anticipation offer.}
    \label{fig:plot_A_as_func_V_c_p_g_s_g}
\end{figure}

As already mentioned, the type of guarantor we are discussing here becomes more relevant the bigger the asymmetry between the default probability estimates: the platform assigns a high default probability to a borrower, while the guarantor estimates a (much) lower value\footnote{The anticipation formula now also includes the three extra parameters $p_g$, $s_g$ and $V_c$.}. This is what is shown in Figure \ref{fig:plot_A_as_func_V_c_p_g_s_g}, where, as an example, the platform default estimate is 10\%, and we consider two situations, one where a guarantor estimates this probability to be 8\% and one where it estimates it to be 4\%. The difference between the anticipation offer with and without guarantor grows with the difference between the two default probability estimates, showing that this kind of guarantor would be most relevant in situations with bigger \emph{information asymmetry}, guarantors being able to make better risk assessments than the platform. As also shown, the bigger the collateral/stake, the better becomes the anticipation offer.

Such situations, where guarantors are better at estimating risks, may occur, for example, in new markets, where data may not be available, or of poor quality. In any case, whenever a guarantor is willing to stake and asks for a \emph{reasonable} extra gain\footnote{Since this gain is discounted from the anticipation, if they ask for too much, it will automatically make the offer unattractive}, even if they happen to be wrong on their risk assessment, the impact for the platform is minimized (\emph{hedged}), given the collateral.

\subsubsection{Investor $\longleftrightarrow$ Liquidity Pool: Providing resources to the system}
\label{sec:investors_liq_pool}

Investors provide resources for the loans/anticipations, expecting an extra spread in return (with respect to the base rate $r$). The greater this spread, the better for investors, as long as there is still demand to efficiently allocate the liquidity pool resources. This balance, in order to attract both, investors and borrowers, is adjusted by the platform via its spread rate $s$. Default rates also impact returns. Even with a perfect rating engine, returns would be as expected only on average, with a certain level of variance. Different investors may have quite different strategies/criteria when making investment decisions (both, to invest and to withdraw). Since this influx of investments is an important component of the simulations, we have to include it. 

In our simulations, investors decide when to invest/withdraw only by considering the funds historical performance, which is a simplified version of the complex reality of investment decision policies. These are the main parameters that describe each investor\footnote{As with other agents, for each investor instance, a set of parameters is sampled from specified distributions.}: (i) investment amount; (ii) (annualized) expected performance over a certain number of periods; (iii) loss and profit withdrawal rate, where, with a certain withdrawal probability, investors withdraw their investments; (iv) minimum holding period.

\begin{figure}[h!]
    \centering
    \includegraphics[width=0.4\textwidth]{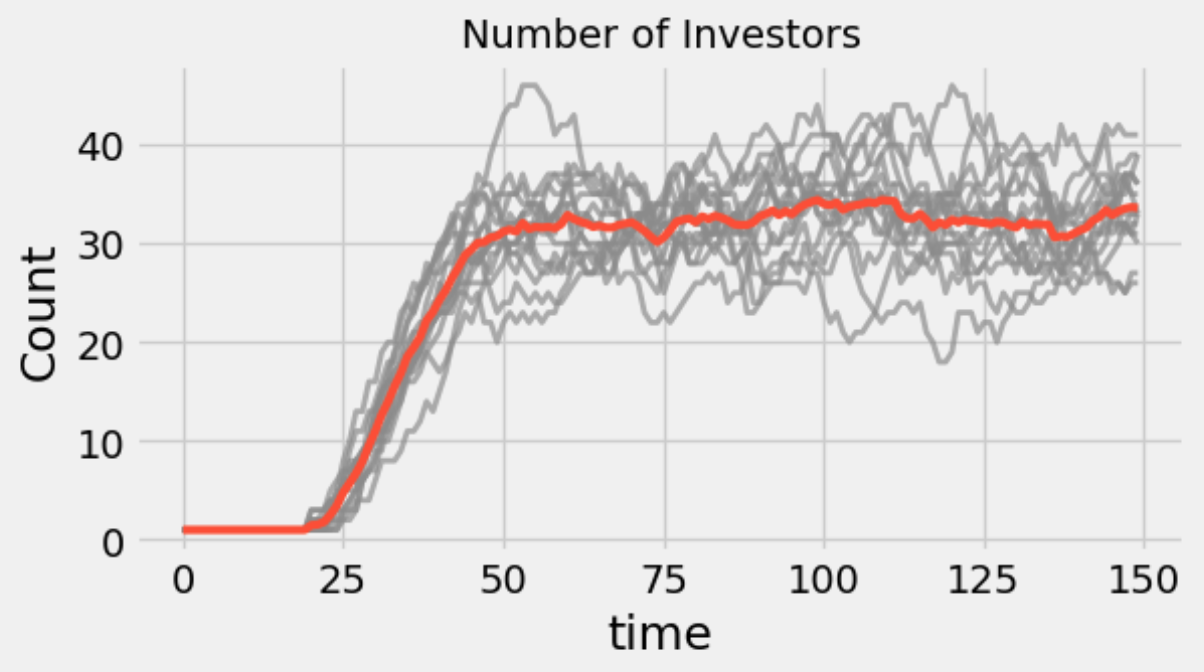}
    \caption{Different simulation outcomes while keeping the input parameters fixed. The gray lines represent the number of investors at each moment of time, for each individual simulations, while the red line is the average from all the simulations.}
    \label{fig:plot_example_num_investors}
\end{figure}

Given the stochastic nature of the simulations, even when all input parameters are held fixed, outcomes will vary. As an example, consider Figure \ref{fig:plot_example_num_investors}, where we show the number of investors over time during different simulations (gray lines), with a fixed set of input parameters. Without getting into the details of why these curves look the way they look -- section \ref{sec:scenarios} provides a more comprehensive discussion of simulations --, we can see the variability between simulations, in this case, the number of investors over time.

The parameters we have just described specify the investors population, and how this population will behave during the time evolution of the system -- even if it is an stochastic behavior. From the borrowers side, the loan demand side, the characterization is a bit more complex, and this is what we discuss on the next section.

\subsection{Borrowers anticipation demand modeling} \label{sec:demand_model}

Given that there are two parties involved on the anticipation \emph{negotiation}, borrowers may or may not accept the platform's anticipation offer. It is important, thus, to determine the \emph{price sensitivity} of borrowing -- i.e. model borrowing demand. Without data to fit a model to infer the price elasticity of credit demand, we'll assume an analytical approach and tune its parameters in order to emulate different price sensitivities for the borrowers. This demand may be influenced by the the presence of guarantors, since the anticipation value may change. However, here we assume that borrowers only consider the anticipation \emph{price}, i.e., they do not distinguish a particular offer having the same price due to the presence of guarantors or because the platform estimated a lower default probability.

\begin{figure}[h!]
    \centering
    \includegraphics[width=0.8\textwidth]{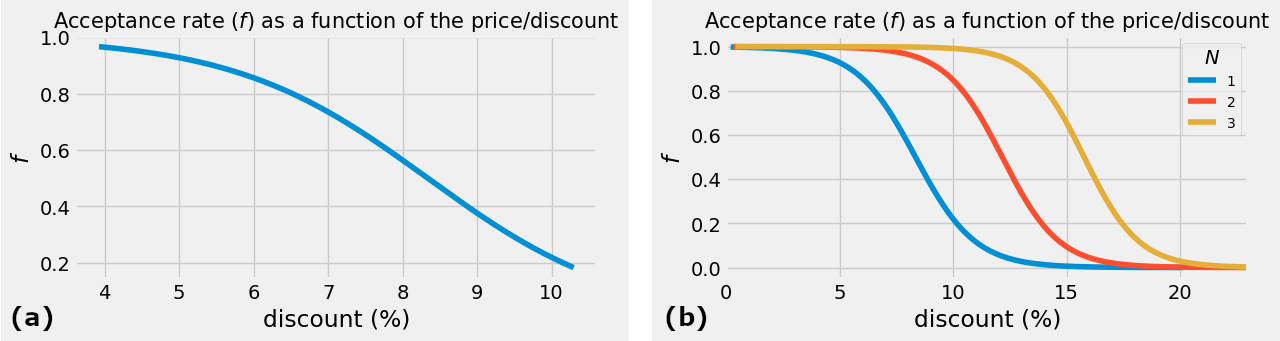}
    \caption{Example of price sensitivity curves as a function of the anticipation discount (a). As made explicit on the right plot (b), these curves depend on parameters like the number of installments $N$.}
    \label{fig:plot_example_f_as_func_discount}
\end{figure}

From the platform's perspective, the main pricing mechanism is setting its spread $s$. From the borrowers' side (the demand side), however, the anticipation \emph{price} is probably the main factor they are sensitive to -- here, by \emph{price}, we mean the discount with respect to the sum of the receivables in negotiation. Also, we'll consider that the borrowers' response to a particular offer is probabilistic, let's call it $f(A)$, where $f(A)$ is a number between $0$ and $1$, representing the acceptance probability for that particular offer. Figure \ref{fig:plot_example_f_as_func_discount}-a shows an example of such a function: as the anticipation discount increases, the acceptance probability decreases.

Given that the anticipation offer depends on a number of factors, such as the number of installments $N$ and the base rate $r$, this price sensitivity curve $f$ must also depend on these factors, and Figure \ref{fig:plot_example_f_as_func_discount}-b shows an example on the dependency on $N$: as the number of installments increases, so does the tolerance for bigger discounts\footnote{There is no unique price reference, it depends on the particular conditions, it could be an \emph{environmental} parameter like $r$, or a \emph{contract} aspect, like $N$.}.

An advantage of this approach is that these curves can be constructed in two ways: (i) by tuning their parameters in order to represent different market conditions, e.g. more/less elastic demand; (ii) by fitting them to data\footnote{Acquiring such data can be challenging, but also, properly fitting such a model to observational data, may also be a difficult task.}. For now, we'll consider only the first approach, as we have not obtained data to fit yet.

\section{Simulations} \label{sec:scenarios}
 
The previous section introduced how we modeled the interactions of our set of agents, borrowers, investors and guarantors  with the DeLend platform. We can now show examples of simulations outcomes.

\begin{figure}[h!]
    \centering
    \includegraphics[width=0.8\textwidth]{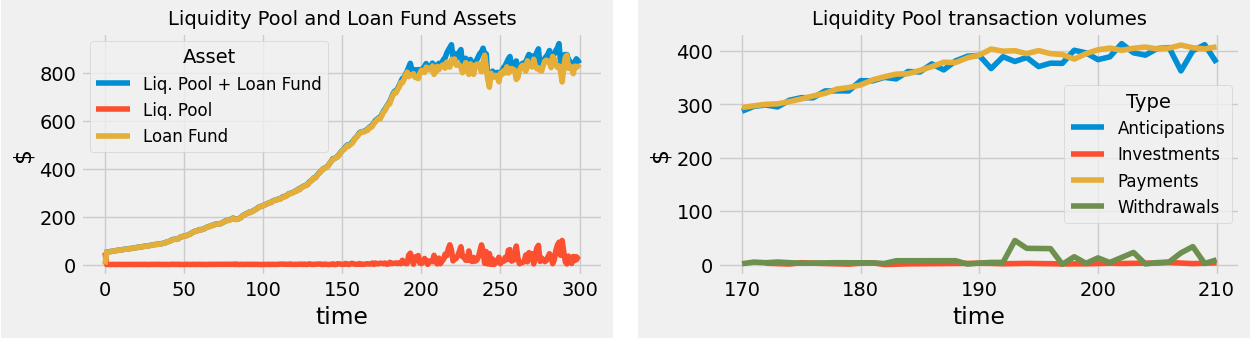}
    \caption{Time evolution of the assets, loan fund and liquidity pool (left). On the right we show the liquidity pool transaction volumes per type of transaction (a subset of the time window for clarity).}
    \label{fig:plot_example_funds_paper}
\end{figure}

Figure \ref{fig:plot_example_funds_paper} (left plot) shows the time evolution of the liquidity pool and of the loan fund, the sum of them compose the total assets resulting from the investments and the loans (see \cite{delendwp1}). In this simulation we have kept the total number of borrowers constant. As a consequence, after a certain number of periods ($\sim190$), the market \emph{saturates}, when loan demand is fully fulfilled. As already mentioned, whenever the allocation of the liquidity pool is sub-optimal, the overall performance is also sub-optimal (lower than expected return rates), and investors start to withdraw (from the liquidity pool, see \ref{fig:plot_example_funds_paper} right plot). This re-balances the overall performance. However, this mechanism is not completely efficient and there are periods with lower allocation of the liquidity pool.

\begin{figure}[h!]
    \centering
    \includegraphics[width=0.8\textwidth]{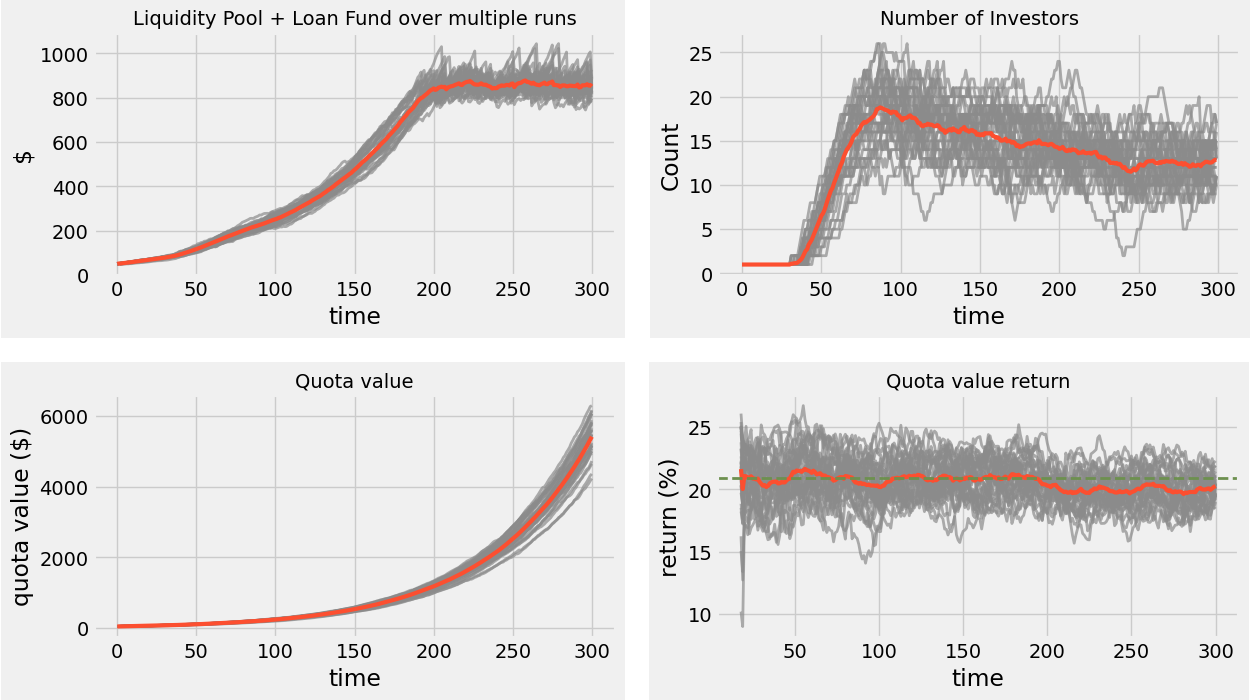}
    \caption{Time evolution of a set of metrics that can be computed for each simulation. The green dashed line at the returns plot is the target return, and returns were computed using an 18 period moving window (bottom right plot).}
    \label{fig:plot_example_metrics}
\end{figure}

A second important outcome from the simulations is to evaluate variances. Even when input parameters are kept constant, each simulation run results, in general, in different outcomes. Different metrics can be computed after each simulation, like the number of investors, the quota value and the historical returns over a certain number of periods (see Figure \ref{fig:plot_example_metrics}). Each metric will have its own average behavior and variance.

\begin{figure}[h!]
    \centering
    \includegraphics[width=0.8\textwidth]{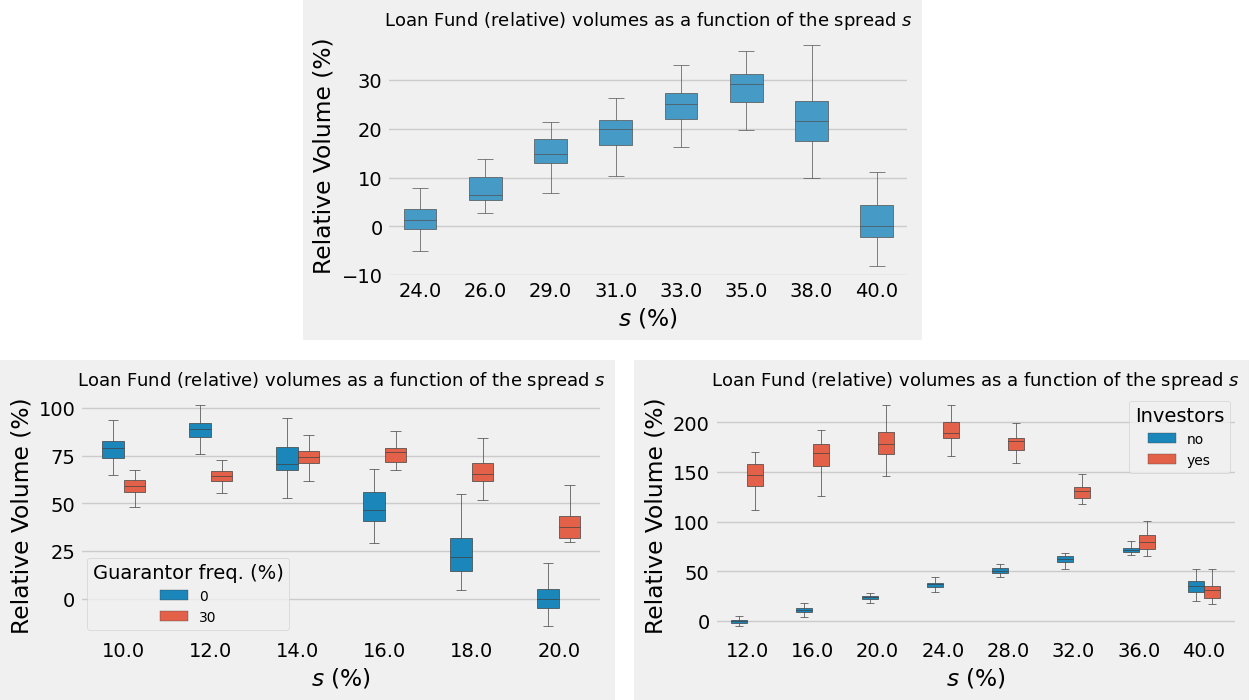}
    \caption{Optimization of the platform's spread rate for different scenarios. Here we use relative volumes, considering the lowest volume at each scenario as the reference value. The results are represented using boxplots to express the variance too.}
    \label{fig:plot_optimization}
\end{figure}

Finally, we can leverage simulations to estimate optimal operational parameters, like the platform's spread $s$. As an example, let's assume we wanted to maximize the loan's fund volume over a period of 36 months, keeping the spread constant during the time evolution of the system, and considering different scenarios. The first scenario considers a situation with just one seed investor (Figure \ref{fig:plot_optimization} top plot). In this case, the optimal spread would be $s \sim 35\%$ (annualized value). The optimal values for the spread come from the relation between \emph{price} (spread) and demand: as the spread increases, the \emph{volume} per loan increases, but, at the same time, demand decreases (see section \ref{sec:demand_model} and Figure \ref{fig:plot_example_f_as_func_discount}). Since the total volume is the product of these conflicting effects, there will always be an optimal spread value -- we may also choose different objective functions, instead of  the loan's fund volume.

Considering the same scenario (input parameters), but now simulating a biased risk assessment engine, one that overestimates the default probabilities, with and without guarantors, we obtain the reusts shown in Figure \ref{fig:plot_optimization} bottom left plot\footnote{The frequency of guarantors in the population is also an input paramter, for this examplo we have set it to be 30\%, meaning that each borrower $\leftrightarrow$ platform interaction has a $0.3$ probability to involve a guarantor.}. The consequence of overestimating the default probability is to overcharge for the loans/anticipations. In order to compensate, the platform's spread must be lowered to obtain optimal results.

To conclude, instead of having a biased risk assessment engine, we introduce investors to the system (Figure \ref{fig:plot_optimization} bottom right plot). As expected, investors can significantly improve loan volumes, especially for cheaper spreads.

\section{Final remarks and next steps}

We have presented our agent-based methodology to be able to simulate the behavior of the DeLend platform. As shown in section \ref{sec:scenarios}, we can consider realistic scenarios, and use the simulations in order to estimate outcomes, depending on factors such as default rates distributions, rating engine capabilities and benefits that guarantors may bring.

The numerical results suggest that the system is capable of governing the channeling of resources from investors to borrowers with a reasonable return for investors and relative stability of the liquidity pool. They also reveal that a good engine for rating loans is essential, and that guarantors are an important tool to improve the performance of the system.

The simulation strategy described in this paper can be extended to richer settings, with parameters adjusted to fit real data and more functionalities added to the system. For instance, other classes of guarantors should be considered\footnote{The system admits different roles for guarantors additionally to those examined here. For instance, multiple guarantors could be simultaneously used for sets of loans, as stakers in mutual funds or mutual guarantors in the joint liability arrangements discussed in \cite{ghatak1999economics}}. The model could also be extended to richer environments, including situations in which the economy is subject to shocks. In this case, it is important to model a dynamic behavior of the platform parameters to respond to such shocks.  

At this point, we have not yet focused on the optimal choice of platform parameters. This is a key numerical exercise to be performed in future work, once a good fit to real data is obtained. 

As next steps, we plan to:

\begin{itemize}
    \item Obtain \emph{real world} data in order to validate/adjust our approach.
    \item Implement a secondary market so that investors can trade their quotas, as a Decentralized Finance (\emph{DeFi}) marketplace. 
    \item Add transaction fees to the model.
    \item Include tax effects, which may depend on specific jurisdictions.
    \item Incorporate different liquid assets in the Liquidity Pool and hedge mechanisms for currency tokens exchange risks. 
    \item As exemplified during the discussion about the platform's spread (section \ref{sec:scenarios}), depending on the scenario, there will be optimal values for the platform's parameters. We plan to perform optimisations based on the simulations in order to evaluate how these parameters could be optimally set for a wider variety of scenarios.
\end{itemize}

\printbibliography

\appendix

\pagebreak

\section{Appendices}

The following set of appendices provide more in-depth discussions regarding: 
\begin{itemize}
    \item \ref{append:anticipation_formula} - Derivation of the anticipation formula without guarantors.
    \item \ref{append:guarantor} - Anticipations involving guarantors, and the anticipation formula in this context.
    \item \ref{append:unit_tests} - A set of \emph{unit tests} for the simulator components, in order to validate them.
\end{itemize}

\subsection{Anticipation formula} \label{append:anticipation_formula}

In order to compute the anticipation offer, we first ignore the possibility of default, and consider only the time discount. Since the borrower will receive the full anticipation amount $A$ immediately in cash, and the receivables $R_i$ will be cashed by the platform in future instalments, these $R_i$ must be discounted by at least a risk free interest rate (a base rate)\footnote{For instance, in Brazil, a reasonable \emph{proxy} for a risk free rate is the CDI, which is the interbank rate set by the Central Bank.}. Denoting this rate as r, in the default-free case, the anticipation value would be the sum of the $R_i$ brought to present value (we can treat each receivable independently, and let’s call it the \emph{base rate anticipation} $A_{br}$)\footnote{The formulas consider a generic number of installments $N$, with $N = 4$ for the example we provided in section \ref{sec:borrower_platform}.},

\begin{equation} \label{eq:A_br}
    A_{br} = \sum_i^{N} \frac{R_i}{(1 +r)^i}
\end{equation}
In order to obtain a better return, the platform may also add a spread $s$, and the \emph{no default anticipation} $A_{nd}$ becomes, 

\begin{equation} \label{eq:A_nd}
    A_{nd} = \sum_i^{N} \frac{R_i}{(1 + r + s)^i}
\end{equation}

This extra (compound) percentage $s$, on top of the base rate, is  the additional return for investors over the risk-free investment options.

In real situations, however, there will always be some level of default. Risk assessment is  a crucial task for the platform, since losses related to default should be considered by investors. A good risk assessment also minimize informational frictions, such as adverse selection (\cite{stiglitz1981credit}). The simulator can consider both, the possibility of default and of delayed payments, but for the simulations discussed here, have ignored the latter. 

\begin{figure}[h!]
    \centering
    \includegraphics[width=0.6\textwidth]{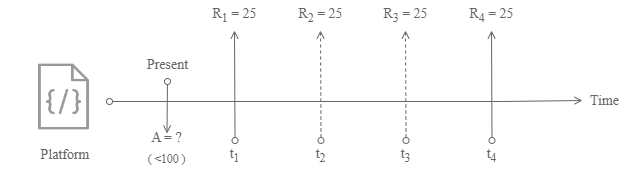}
    \caption{Cash flow where $R_2$ and $R_3$ are not paid on time}
    \label{fig:baisc_inst_cflow_def}
\end{figure}

\begin{figure}[h!]
    \centering
    \includegraphics[width=0.6\textwidth]{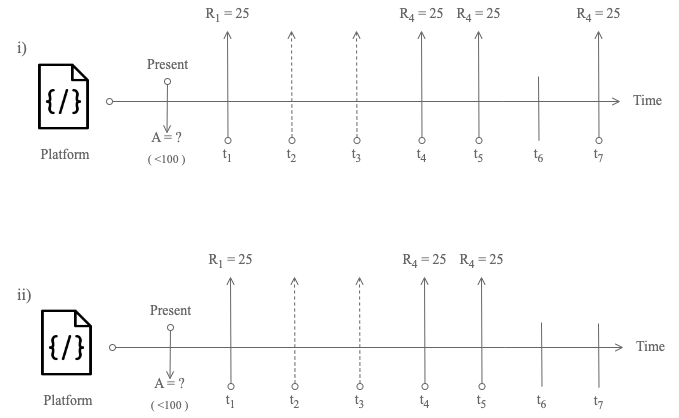}
    \caption{Cash flow with delays (\emph{i}), and with both, delays and defaults (\emph{ii})}
    \label{fig:basic_inst_cflow_def_outcomes}
\end{figure}

As an example, suppose two of the four installments ($R_2$ and $R_3$, dashed arrows) are not honored in their due time, as shown in Figure \ref{fig:baisc_inst_cflow_def}. In this scenario, there are two possible outcomes (Figure \ref{fig:basic_inst_cflow_def_outcomes}): \emph{i)} all the installments/receivables are eventually paid; or \emph{ii)} payments are only partially honored.

Figure \ref{fig:basic_inst_cflow_def_outcomes} depicts a particular view, where we consider the second payment to be $R_2$, irrespective of the period, the third to be $R_3$, and so on. One advantage of this formulation is that there will not be ambiguities to classify any payment. 

In the numerical exercises presented in this document, for simplicity, we focus on a subset of the possible profiles of payment. \textbf{We define the default probability $p$ to be the probability that a borrower expected to pay an installment at a given period stops making payments}. We can certainly model the payment behavior in more elaborate ways, but for now we’ll consider this simple definition. A few consequences of this definition are:

\begin{itemize}
    \item Payments have (are drawn from) a delay probability distribution.
    \item Borrowers who have not yet defaulted may default at any period, with constant probability. Once they fail to make a payment, no further installments are paid. In this sense, this default probability can be interpreted as a bankruptcy probability.
\end{itemize}
These definitions are important because they determine the computation of anticipation offers, as we shall discuss now.

If we ignore delays, $p_{d,i}$, the probability that a borrower does default at time $t_i$, is given by the geometric probability distribution,

\begin{equation}
    p_{d,i} = (1 - p)^{(i - 1)}p
\end{equation}
We can interpret this probability as being composed by two terms: \emph{i)} $(1-p)^{(i-1)}$ is the probability that the borrower does not default on the first $(i-1)$ periods; and \emph{ii)} $p$ the (constant over time) probability that the borrower ends up defaulting at i. The probability $ p_{\bar{d},i}$ that the borrower does not default at time $t_i$ is,

\begin{equation}
    p_{\bar{d},i} = (1 - p)^{(i - 1)}(1 - p) = (1 - p)^{i}
\end{equation}
One important consequence of this formula is that the default probability decreases as a function of the time period, not because defaults become less likely, but rather due to the fact that borrowers’ payments may cease in earlier periods: the further in time, the smaller the number of borrowers that have not yet defaulted. 

We need these probabilities in order to compute the expected losses/payments associated to a set of receivables to be paid by a borrower for which we attribute a default probability p. The expected payment for each installment at $t_i$, $R’_i$, is the probability that the borrower does not default at time $t_i$ times the installment amount $R_i$,

\begin{equation}
    R’_i = (1 - p)^i R_i
\end{equation}
If we use the expected payment in the anticipation price formula, then \emph{final} anticipation formula (not yet including guarantors) becomes,
\begin{equation} \label{eq:A_no_g_final}
    A = \sum_i^{N} \frac{R'_i}{(1 + r + s)^i} = \sum_i^{N} \frac{(1 - p)^i R_i}{(1 + r + s)^i}
\end{equation}

\subsection{Guarantors} \label{append:guarantor}

 The interaction/negotiation between the platform and a guarantor can be modeled in different ways. Here, we consider a situation where guarantors have to stake an amount $V_c$ (a collateral that is lost in case of borrower default) in return of a gain $G$. A simple approach, which is advantageous from a computational perspective, is to consider that the guarantor sets its desired gain $G$ and communicates it to the platform. The platform then decides (computes) if it considers it advantageous, and makes the anticipation offer (to the borrower) including or excluding the guarantor. The guarantor computes its desired gain as follows.

\begin{figure}[htp]
    \centering
    \includegraphics[width=0.5\textwidth]{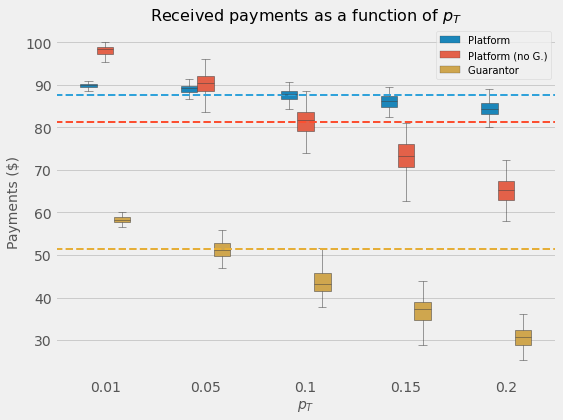}
    \caption{Asked gain $G_s$ as a function of $p_g$ and $s_g$, with $r = 1\%$ (annualized).}
    \label{fig:plot_gain_sg_pg}
\end{figure}

The guarantor has an amount $V_c$ to stake as a collateral, an estimate of the default probability $p_g$, knows the base rate is $r$ and has an expected spread $s_g$. The platform (in principle) does not know the guarantor's default estimate nor $s_g$. Since the guarantor looses the collateral in case of default, its expected gain will be estimated as (considering an arbitrary number of $N$ installments),

\begin{equation}
    E[G] = (1 - p_g)^N \left( V_c(1 + r)^N + G_s\right)
\end{equation}
where $G_s$ is the extra gain to be paid by the platform. Given the guarantor's desired spread is $s_g$, the guarantor will stipulate its (expected) gain to be,

\begin{equation}
    E[G] = V_c(1 + r + s_g)^N
\end{equation}
These relations define $G_s$ to be,

\begin{equation}
    G_s = V_c(1 + r + s_g)^N/(1 - p_g)^N - V_c(1 + r)^N
\end{equation}
Note that the formula above incorporates the fact that for higher (estimated) risks $p_g$, guarantors ask for higher returns (Figure \ref{fig:plot_gain_sg_pg}). Considering this expected gain, the platform computes the anticipation offer and checks if it is feasible -- in the sense of being advantageous for all the players. We now show how the platform computes the anticipation offer considering the guarantor's stake and requested gain.

The platform must consider these three cash flows: (i) the borrower's receivables/installments; (ii) the gain $G_s$ to be paid to the guarantor once the last installment arrives; and (iii) the collateral, that the platform gets to keep in case of default and that must be adjusted by the risk free rate $r$. If, for a moment, we ignore the base rate and the platform's spread (set both to zero), and write the anticipation value as a sum these cash flows expected values:  (i) $E[R]$, the receivables expected value; (ii) $E[G_s]$ the expected value to be paid to the guarantor -- this amount is discounted from the anticipation, thus, it is \textit{paid} by the borrower; and (iii) $E[L_c]$ the guarantor's collateral expected loss; we have, calling it the \emph{no spread} anticipation $A_g^{no-spread}$,

\begin{equation}
  A_g^{no-spread}  = E[R] - E[G_s] + E[L_c]  
\end{equation}
where,

\begin{equation}
    E[R] = \sum_i^N (1 - p)^iR_i
\end{equation}

\begin{equation}
    E[G_s] =  (1 - p)^NG_s
\end{equation}

\begin{equation}
    E[L_c] = \sum_i^N (1 - p)^{(i -1)}p V_c(1+r)^i
\end{equation}
It is important to note the while the guarantor computes expected values using its default estimate $p_g$, the platform considers its own estimate $p$.

The above anticipation $A_g^{no-spread}$ corresponds to the value the platform should anticipate in order to, on average, cover the default losses, with no profitability -- and when $r=0, s=0$. However, since the base rate will not be zero, and, to deliver a higher return, we must add a spread $s$, we have to insert back the discounts $(1 + r +s)$ to each of these cash fluxes, and the anticipation offer formula becomes, 

\begin{equation} \label{eq:A_g}
  A_g = \sum_i^N \frac{(1 - p)^i\left(R_i - \delta_{i,N}G_{s}\right)  + (1 - p)^{(i -1)}p V_{c,i}}{(1 + r + s)^i}  
\end{equation}
where we used the notation,
\begin{equation}
    V_{c,i} = V_c(1 +r)^i
\end{equation}
\begin{equation}
    \delta_{i, N} = 
    \begin{cases}
        \ 1, & \text{if } i=N \\
        \ 0, & \text{otherwise}
    \end{cases}
\end{equation}

\begin{figure}[htp]
    \centering
    \includegraphics[width=.8\textwidth]{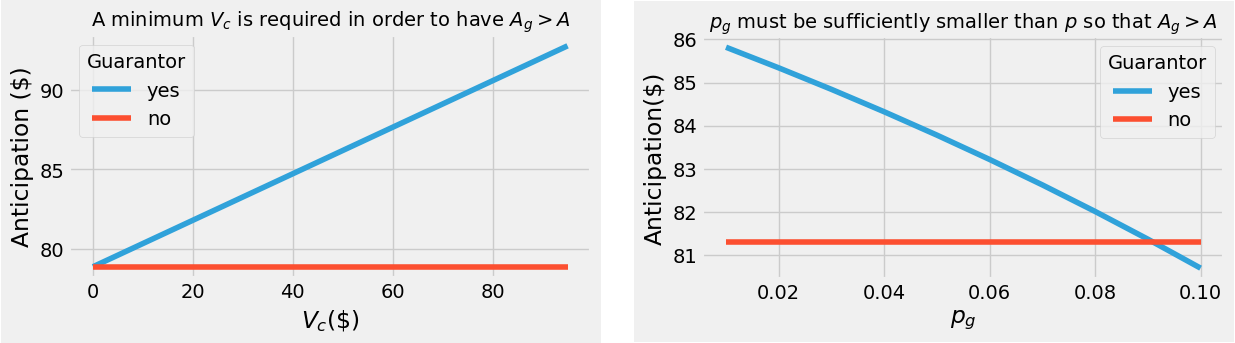}
    \caption{Anticipation, where the receivables sum to 100: as a function of $V_c$, with $p=0.1$,$p_g=0.05$, $s_g=0.01$ and $N=3$ (left); and as a function of $p_g$, with $p=0.1$, $V_c=20$, $s_g=0.01$ and $N=3$ (right)}
    \label{fig:plot_ant_vc_pg}
\end{figure}

From equation (\ref{eq:A_g}), we can see that:
\begin{itemize}
    \item The larger the collateral $V_c$, the better the anticipation, which, at least indirectly, is an incentive for guarantors to stake bigger amounts -- since it increases the borrowers' acceptance probability. As shown in Figure \ref{fig:plot_ant_vc_pg} (left), as $V_c$ becomes smaller the anticipation tends to become the same as without guarantors.

    \item As the guarantor's estimated default probability $p_g$ decreases, so does its asked gain $G_s$ -- which increases the anticipation value. At the same time, if the platform's estimated default $p$ increases, the discount to the anticipation coming from $G_s$ gets smaller -- since the the gain is only paid in the borrower does not default, and the anticipation considers expected values. These two factors combined lead to the guarantor's involvement being advantageous when 
    $p_g < p$ (Figure \ref{fig:plot_ant_vc_pg}, right).
\end{itemize}

\begin{figure}[h!]
    \centering
    \includegraphics[width=0.5\textwidth]{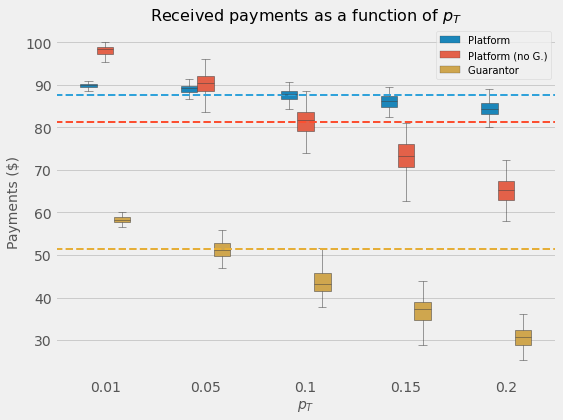}
    \caption{Simulation of received payments where $N=3$ (three installments that sum 100), $p=0.1$, $V_c=50$, $p_g=0.05$, and $s_g=0.02$, as a function of the true default probability $p_T$ (here, again, $r=0$ and $s=0$). The data represented by the boxplots was generated considering 100 borrowers, and their payment behaviour was repeated 100 times. At each iteration the average payment was computed. These average payments are represented by the boxplots. Note that the blue and red lines coincide with the received payments when $p_T = 0.1$, which is the platform's default probability estimate, and the yellow line coincide with the guarantor's received payment when $p_T=0.05$, the guarantor's default probability estimate.}
    \label{fig:plot_sim_pT}
\end{figure}

Finally, although, in principle, guarantors are better informed agents, thus having a better risk assessment ($p_g$ more accurate than $p$), the true default probability $p_T$ is not known. Figure \ref{fig:plot_sim_pT} shows the results of a simulation with the payments received by \emph{i)} the platform (with guarantors); \emph{ii)} the platform without guarantors; and \emph{iii)} the guarantors. The dashed lines correspond to: \emph{i)} the anticipation value with guarantors (blue line); \emph{ii)} the anticipation value without guarantors (red line); and \emph{iii)} the guarantors expected payment (yellow line). 

As expected, whenever the true probability is bigger than the estimated by the agent (platform or guarantor), the received payments are less than the expected amounts: the platform expects to receive $\sim 87.5$ when guarantors are involved, $\sim 81.3$ when they are not involved, and the guarantors expect a payment of $\sim 51.5$. However, guarantors also have a \emph{hedging} effect for the platform: the difference between the expected payment (blue dashed line) and the actual received amounts (blue boxes) are considerably smaller than when guarantors are not involved (red line/boxes). This effect is, of course, dependent on the amount $V_c$ staked by the guarantor, since, as this amount decreases, the results would converge to the situation without guarantors.

\begin{figure}[h!]
    \centering
    \includegraphics[width=0.5\textwidth]{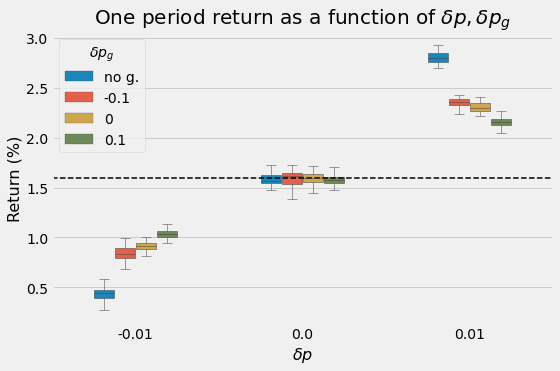}
    \caption{Simulation of average one period returns, where the true default probability if 0.15. As for the platform and guarantors' estimates, they are measured with respect to the true default probability, by $\delta p$ and $\delta p_g$, respectively -- for example, $\delta p = 0.1$ means the platform's estimated probability is $p = 0.16$. The dashed dashed black line represents the expected return $r + s$. The situation where guarantors are not considered are also included (labeled as \emph{no g.}, in blue). We can see that: (i) returns are much more sensitive to errors on the platform's probability estimate; and (ii) as already noted above, guarantors also constitute a hedging mechanism for the platform.}
    \label{fig:plot_sim_pT_pg}
\end{figure}

\subsection{Python implementation and simulator \emph{unit tests}} \label{append:unit_tests}

Since agent-based simulations is a well established approach, employed in a wide variety of situations, many existing frameworks exist and can be used to leverage the development of a code to address a particular problem. We have used the \href{https://mesa.readthedocs.io/en/stable}{Mesa framework}, given its adoption, stability and functionalities.

The purpose of this section is to introduce the different components and agents in simplified examples, and these examples also are meant to validate the simulator (kind of \emph{unit tests} for the simulator).

\begin{figure}[h!]
    \centering
    \includegraphics[width=0.8\textwidth]{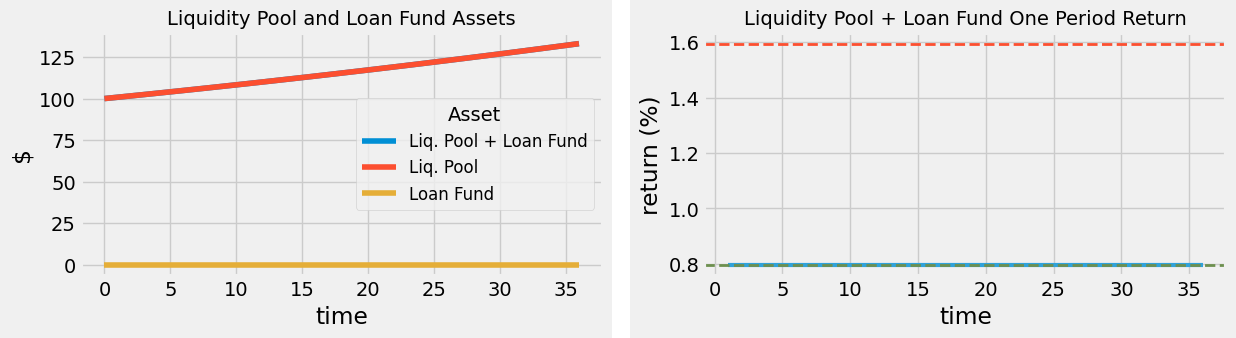}
    \caption{Liquidity Pool, Loan Fund and Total assets time evolution, with no borrowers. Here, the blue and red curves overlap, since there are no loans (left). Time evolution of one period return (Av. Funds + Debts) with no borrowers. The red line corresponds to $r=0.797\%$, $10\%$ per year (right).}
    \label{fig:results_no_borrowers}
\end{figure}

Any dynamics involving loans/anticipations can only begin once there are resources on the liquidity pool. One of the main input parameters (of the simulations) sets the base rate $r$, that affects how much the available resources (at the liquidity pool) are adjusted at each time period. Figure \ref{fig:results_no_borrowers} (left) shows the evolution of the funds: the liquidity pool (available borrowing funds); the loan fund; and their sum. Since there are no loans, the loan fund remains empty in this case.

In order to validate whether the funds evolve as expected, we compute the one period returns for the total assets. Later, once investors are introduced, we'll consider quota returns instead. Figure \ref{fig:results_no_borrowers} (right) shows that in this case the one period return (blue line) is exactly the base rate $r$ (green, dashed line) -- the red dashed line corresponding to $r + s$, which is the targeted return, but without borrowers it is not achieved.

\begin{figure}[h!]
    \centering
    \includegraphics[width=0.5\textwidth]{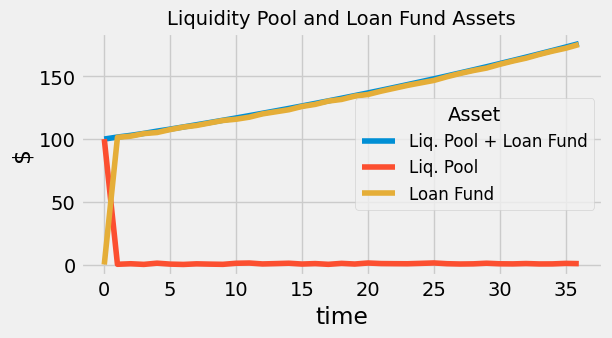}
    \caption{Liquidity pool assets time evolution with (almost) full allocation and no default. At the end of 36 periods, Av. Funds + PV Debts = 176.1, whereas $(1 + r +s)^{36} = 1.768$. This (small) difference is due to the not 100\% allocation of the liquidity pool over time. Note that time $t = 0$ corresponds to the initial configuration, prior to any borrowing.}
    \label{fig:plot_liq_pool_assets_basic_no_default}
\end{figure}

\begin{figure}[h!]
    \centering
    \includegraphics[width=0.5\textwidth]{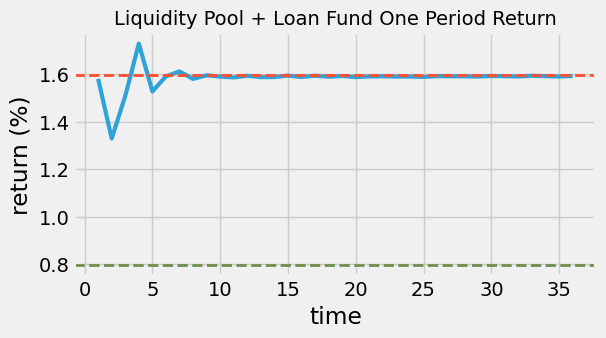}
    \caption{Time evolution of one period returns (Av. Funds + Debts) with (almost) full allocation and no default. The red line corresponds to $r + s=1.59\%$ ($10\%$ per year each). Initially, there is a transient phase but then the one period return tends to converge to $r + s$ -- as mentioned above, since the Liquidity Pool's allocation is never 100\%, the return is slightly smaller.}
    \label{fig:plot_liq_pool_no_default_one_period_return}
\end{figure}

Next, borrowers are introduced, at first with 0 default probability. The total assets returns must now be $r + s$, with (very close to) optimal allocation of the liquidity pool (resources continuously consumed by loans), and this is what we obtain (Figures \ref{fig:plot_liq_pool_assets_basic_no_default} and \ref{fig:plot_liq_pool_no_default_one_period_return}). 

It is reasonable to assume a market much bigger than the platform's lending resources, in the sense that we would never reach a \emph{saturated market} situation, with loan demand entirely fulfilled\footnote{Fulfilled in the sense that they are either accepted, or rejected by the borrower or the platform -- and not unfulfilled due to unavailable resources.}. However, we do want to test this scenario, of saturated market, and validate that the simulations reproduce the expected results.

\begin{figure}[h!]
    \centering
    \includegraphics[width=0.5\textwidth]{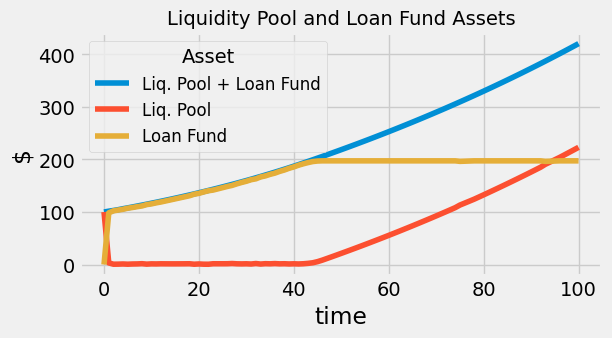}
    \caption{Liquidity pool assets time evolution with a limited number of borrowers. At some point, there is enough credit to fulfill the \emph{market} demand -- the borrowed amount/debt becomes constant. At this point, the return starts to decrease, since the liquidity pool's allocation becomes more and more inefficient (see Figure \ref{fig:plot_liq_pool_no_default_saturated_one_period_return}).}
    \label{fig:plot_liq_pool_assets_basic_no_default_saturated}
\end{figure}

\begin{figure}[h!]
    \centering
    \includegraphics[width=0.5\textwidth]{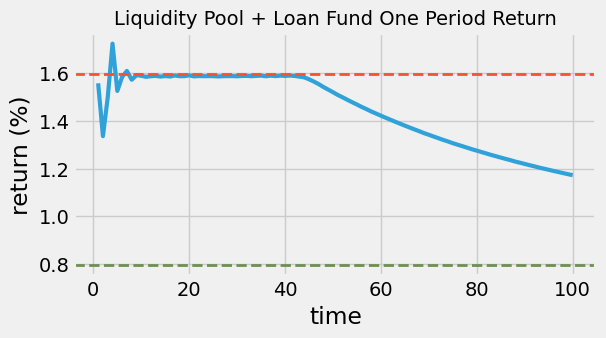}
    \caption{Time evolution of one period return (Av. Funds + Debts) with a limited number of borrowers. The red line corresponds to $r + s=1.59\%$ ($10\%$ per year each) and the green line corresponds to $r = 0.797\%$, the base rate. As the Liquidity's Pool allocation becomes more inefficient, the return starts to converge to the base rate.}
    \label{fig:plot_liq_pool_no_default_saturated_one_period_return}
\end{figure}

In case the excess resources of the liquidity pool are neither consumed by loans/anticipations nor by withdrawals, once the loans reach a maximum constant volume (borrowing demand totally fulfilled), the total assets returns start to converge to the base rate $r$, as the liquidity pool's stalled resources increase, as shown in Figures \ref{fig:plot_liq_pool_assets_basic_no_default_saturated} and \ref{fig:plot_liq_pool_no_default_saturated_one_period_return}.

\begin{figure}[h!]
    \centering
    \includegraphics[width=.8\textwidth]{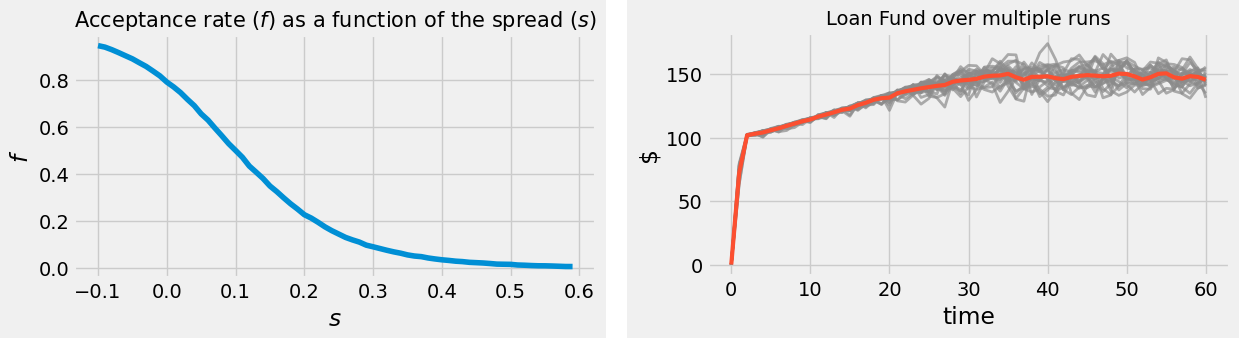}
    \caption{Left, example of an acceptance rate function $f(A)$, and the time evolution of the loan fund over multiple runs (right).}
    \label{fig:plot_example_loan_fund_f_function}
\end{figure}

For the previous example, we have set the parameters such that borrowers always accepted the platform's anticipation offers. However, we can expect them to be price (discount) sensitive, in the sense that as the anticipation offers discounts become bigger (more expensive), the fraction of borrowers willing to accept the offer decreases. As discussed in section \ref{sec:demand_model}, we have modeled this behavior via a function $f(A)$ that assigns an acceptance probability as a function of the anticipation offer $A$. Figure \ref{fig:plot_example_loan_fund_f_function} shows the time evolution of the loan fund (right plot), for a particular parameterization of the $f(A)$ function (left plot).

\begin{figure}[h!]
    \centering
    \includegraphics[width=.8\textwidth]{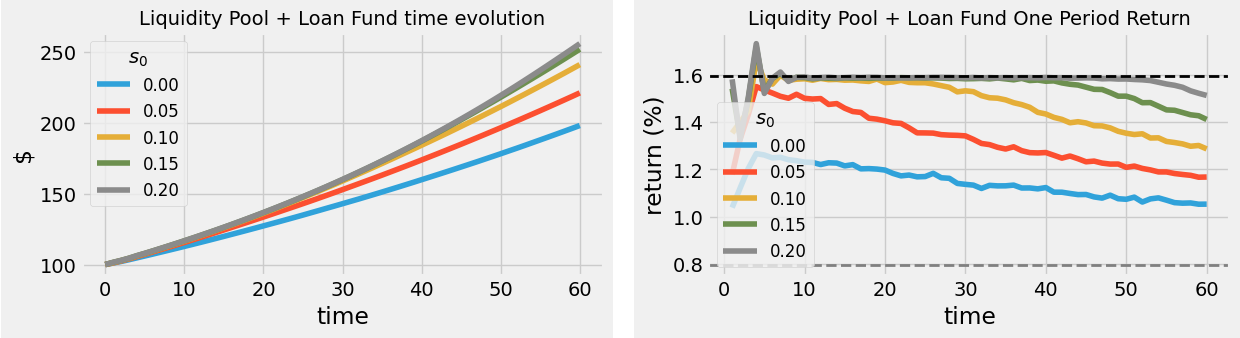}
    \caption{Liquidity Pool + Loan fund (mean) performance for different levels of borrowers acceptance rates (left). (Mean) One period returns for each of the borrowers' threshold spread (right).}
    \label{fig:plot_borrowers_spread}
\end{figure}

The acceptance rate function (a.k.a price sensitivity curve) is defined as follows,
\begin{equation}
    f(A) = \frac{1}{1 + e^{\phi(A - A_0)/A_0}}
\end{equation}
where $A_0$ is the anticipation offer without guarantors and considering a \emph{reference} spread $s_0$: this function is thus parameterized by two parameters, $\phi$ and $s_0$, where the first determines the \emph{spread} of the price sensitivity and the latter the \emph{reference level} (\emph{position}). As $s_0$ decreases, smaller fractions of the borrowers' population accept the anticipation offers (while keeping all other factors constant), and, as a consequence, the platform's capability to obtain funds decreases (Figure \ref{fig:plot_borrowers_spread}).

\begin{figure}[h!]
    \centering
    \includegraphics[width=0.8\textwidth]{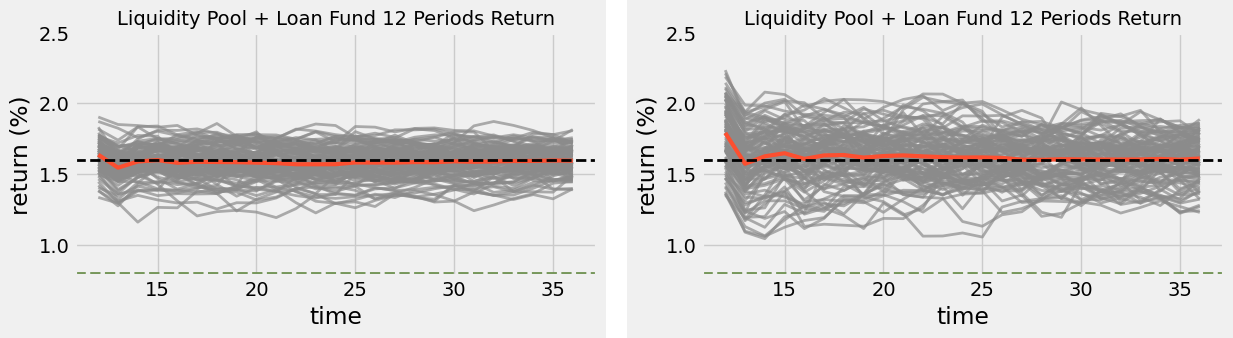}
    \caption {One period returns (100 simulations) for defaults sampled from a Beta(2, 400) (left); and from a Beta (2, 150) (right). The red lines are the averages, and both are very close to the targeted $(r +s)$ return. As expected, the bigger the default, the bigger the variance.}
    \label{fig:plot_one_period_return_default_non_recurrent}
\end{figure}

An important element of the simulations is the borrowers' default probability distribution. We sample, for each borrower, a default probability from a Beta distribution, $p \sim \mathcal{B}(a,b)$, where, for each simulation, we set the values of $a$ and $b$ -- as usual, it is not only a matter of the average value, but also of variance. Figure \ref{fig:plot_one_period_return_default_non_recurrent} compares the outcomes of two different choices for $a$ and $b$. The average outcome is the same, but the variance is different.

Given a perfect \emph{rating} engine, the platform would, on average, obtain the targeted returns. Unfortunately, such an engine does not exist, and we must try to obtain the best balance between bias and variance given the (training) data available. In order to emulate different \emph{qualities} for the rating engine, in a simple way, we have implemented a linear relation between the true default probability and the platform's estimate. Considering $p$ to be the platform's estimate, and $p_T$ to be the true default probability, we have

\begin{equation} \label{eq:rating_linear_model}
    p = ap_T + b + \sigma,
\end{equation}
where $a$ and $b$ are parameters we define on the input for each simulation, and $\sigma$ is sampled from a Gaussian distribution with zero mean and a standard deviation also defined as an input parameter -- a \emph{perfect} engine means $p = p_T$.

\begin{figure}[h!]
    \centering
    \includegraphics[width=0.8\textwidth]{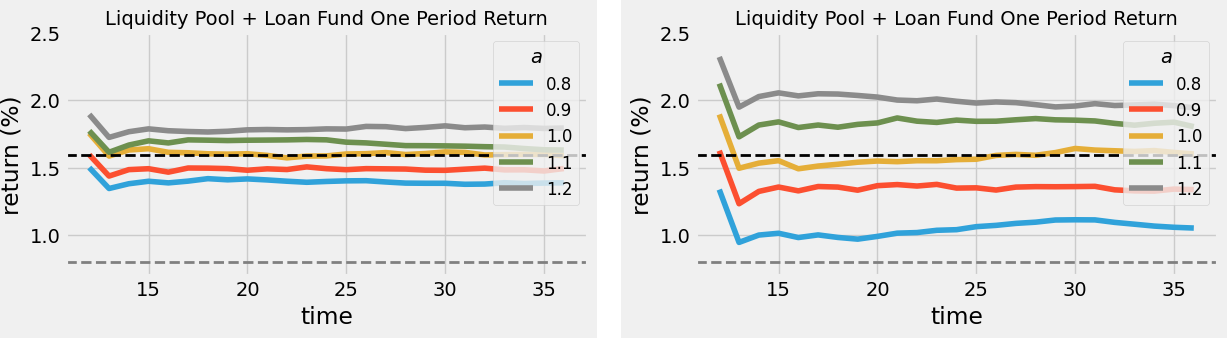}
    \caption {One period returns  as a function of different biases of the rating engine (different values for $a$, see equation \ref{eq:rating_linear_model}), and for defaults sampled from a Beta(2, 200) (left), and from a Beta(2, 80) (right). Note the difference in scales between the two plots: the bigger the default rates, the bigger the impact a not well calibrated rating will have -- potentially becoming negative.}
    \label{fig:plot_average_performance_default_risk_assessment}
\end{figure}

As example, we have simulated (Figure \ref{fig:plot_average_performance_default_risk_assessment}) the impact on the returns using different values for $a$ (from 0.8 until 1.2), while keeping $b$ and $\sigma$ equal to zero, for two different default distributions, one emulating a scenario of relatively low default rates (Beta(10, 1000)), and one of relatively high default rates (Beta(10,100)). While for the low default rates scenario, a less accurate rating engine may still be tolerable, as the default rates increase, without having a good rating, the results quickly start to deviate significantly from the targets -- see Figure \ref{fig:plot_average_performance_default_risk_assessment}, in particular the case where $a = 0.8$ when defaults are high, where the returns are negative, since the engine is systematically (significantly) underestimating the true default probabilities. Even though the goal will always be to have a proper rating engine, there may be situations where where we cannot. For example, when starting in a new market, with limited to no data, it will be wiser to be more conservative with the ratings, assuming \emph{high(er)} default probabilities. This would ensure safer loans, but would also limit the volume of borrowers willing to accept the offers. This is one scenario where guarantors may play an important role, by being able to better assess the borrowers risks.

As discussed in section \ref{sec:guarantors}, guarantors may play different roles, but considering the type we have considered so far, they will only be relevant when there is a significant asymmetry between their default probability assessments ($p_g$) and the platform's assessment ($p$): (i) when  $p_g \sim p$ or $p_g > p$ the anticipation offer when considering guarantors will be worse than without guarantors -- the platform may still consider making an offer involving guarantors due to the \emph{hedging} effect shown in Figure \ref{fig:plot_sim_pT}; (ii) it's when $p_g < p$  that involving the guarantor will tend to become advantageous for all players (the bigger the difference in probabilities, the better, in this sense), since the anticipation offer with guarantors will tend to be better than without involving them -- which makes sense, since the platform would be \emph{overpricing} due to its overestimation of the borrower's default probability.

\begin{figure}[h!]
    \centering
    \includegraphics[width=0.8\textwidth]{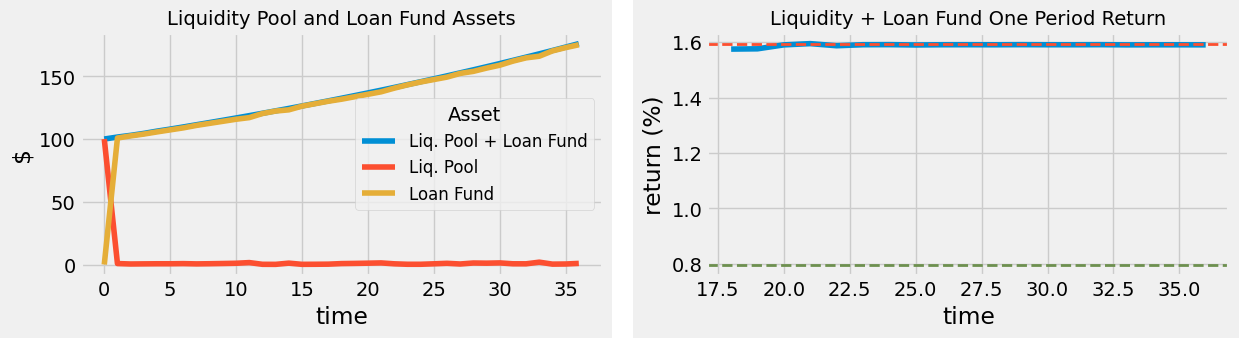}
    \caption {Liquidity Pool + Loan fund performance, where all loans involved a guarantor, with no defaults (left). After a transient initial phase, the average returns converge to the expected values of $r + s$, red dashed line, the green line being the base rate $r$ (right).}
    \label{fig:plot_guarantor_no_default}
\end{figure}

In order to test the guarantors agents, we have first considered a situation where borrowers do not default ($p_T = 0$), with a perfect rating model. In this case, the anticipation offers involving guarantors will actually be lower that without guarantors (but we can make the platform consider this option via input parameters). An important input parameter is the fraction of borrowers that have a guarantor, and in this case we have set it to one. As expected, in this case the system evolves in the same way as if guarantors had not been involved (Figure \ref{fig:plot_guarantor_no_default}).

\begin{figure}[h!]
    \centering
    \includegraphics[width=0.8\textwidth]{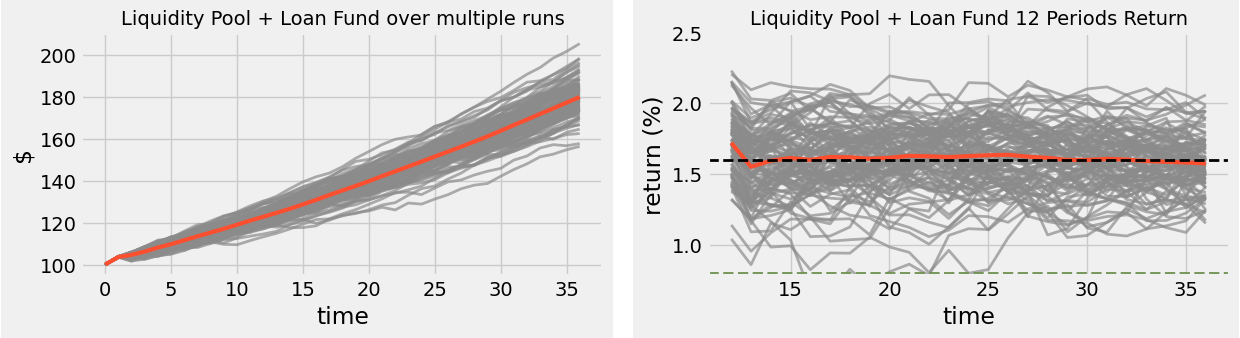}
    \caption {Liquidity Pool + Loan fund performance (100 simulations), where all loans involved a guarantor, for defaults sampled from a Beta(2, 200). After a transient initial phase, the average returns converge to the expected values of $r + s$, red dashed line, the green line being the base rate $r$ (right).}
    \label{fig:plor_guarantor_beta_1770_10000}
\end{figure}

Whenever guarantors are involved, the cash flow management is different (section \ref{sec:guarantors}): stakes must be either re-paid with interests or kept in case of default, and gains must be paid when loans are honored. When there is no asymmetry between $p$ and $p_g$ (the platform's and the guarantor's default estimates, respectively), the platform should obtain the same returns as it would without involving guarantors. This is what we validated (Figure \ref{fig:plor_guarantor_beta_1770_10000}) by considering borrowers with default probabilities sampled from a Beta(2, 200) ($p \sim 1\%$).

The last agent we'll introduce are investors. Their relevance is obvious: they provide  more resources to the system. A set of parameters are used to characterize them -- as with the other agents, most parameters are sampled from distributions specified by input options, in such a way that each investor will have different characteristics/behavior. Amongst the most relevant parameters are: (i) Expected return: the minimum performance in order to invest; (ii) Investment amount: how much to invest; (iii) Profit withdrawal rate: rate that might lead to a withdraw (may be random); (iv) Loss withdrawal rate (again, may be random).

\begin{figure}[h!]
    \centering
    \includegraphics[width=.8\textwidth]{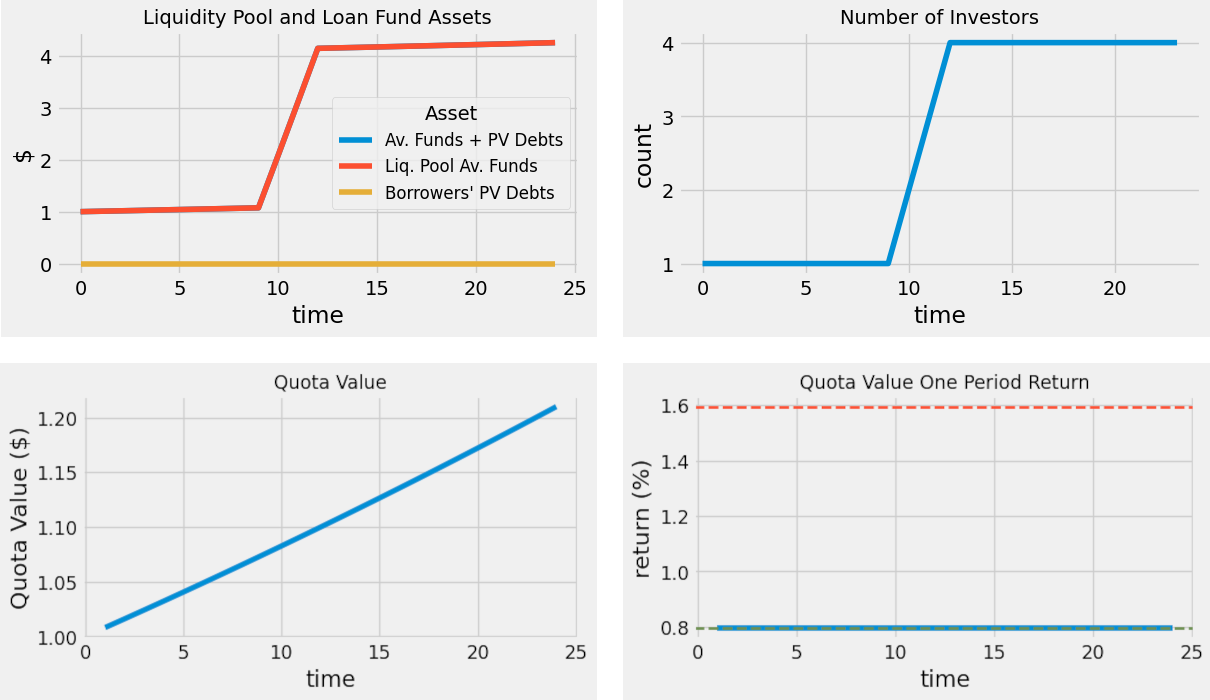}
    \caption {Outcome of a simulation that starts with a seed investor (in this case, the seed investor does not withdraw), followed by a constant flux of investors (one new investor per period) that withdraw after 3 periods, thus stabilizing at a total of 4 investors (top, right) -- for simplicity we did not include borrowers. Once we have this flux of investors, it is necessary to work with quotas (bottom, right), to compute the returns. The bottom left plot shows that the quota returns are indeed the base rate $r$, the dashed green line.}
    \label{fig:plot_invesrors_no_borrowers}
\end{figure}

So far we could compute profits directly from the assets (Liquidity Pool + Loan Fund), however, once we have a flux of investors investing and withdrawing, we need to work with quotas, in order to compute the returns, so from now on, that is how we compute returns. Figure \ref{fig:plot_invesrors_no_borrowers} exemplifies the difference between the quota and the assets. In this simplified example, without borrowers, the return should be the base rate $r$.

\end{document}